# A COSMOLOGICAL MODEL FOR THE EUCLIDEAN 5-SPHERE ARCHAIC UNIVERSE

I. Licata[1], L.Chiatti[2]

April, 2010

## Abstract

In this work, we examine in depth the cosmological physical aspects of the archaic universe described by Euclidean 5-sphere geometry. (see Int. Jour. of Theor., Phys 2009, 48:1003-1018). We hypothesize that the big bang consisted of a spatially extended nucleation process which took place at the end of a pre-cosmic phase, characterized by the evolution parameter $\underline{x}_0$. This parameter, which can be considered a quantum precursor of ordinary physical time, is a coordinate of Euclidean 5-sphere metrics. We must now examine what the privileged role of the $\underline{x}_0$ axis consists in. The big bang is a sort of nucleation of matter and fields by vacuum; to try to understand it we must therefore make some assumptions regarding a pre-cosmic state of matter and energy. The introduction of an evolution parameter such as $\underline{x}_0$ which can be extended to pre-big bang situations is absolutely necessary if we are to define any pre-cosmic dynamics. A generalized Bekenstein relation is here proposed for archaic Universe.

A complete solution to Projective General Relativity (PGR) equations in the De Sitter Universe is provided, so as to establish univocal relations between the scale factor $R(\tau)$ and cosmic time $\tau$. In this way, the physics and geometry of the cosmological model are specified completely.

## 1. Introduction

In a previous article [1] we introduced and justified the concept of "Archaic Universe", which we identified in Arcidiacono's Euclidean 5-sphere. In that context, it had been shown how a spatial but timeless reality (the 5-sphere) can constitute the substratum for the customary spacetime coordinates and their metrics. These ones describe the propagation of wavefunctions between two successive R processes[3], while the true fundamental geometric structure is represented by the surface of the Euclidean 5-sphere.

Furthermore, in the interpretation suggested in [1], the time variable which appears as an argument of the wavefunctions is not a generic time measured by an unspecified local clock; it is rather *cosmic*

---

[1] Institute for Scientific Methodology, Palermo (Italy), e.mail: ignazio.licata@ejtp.info
[2] AUSL VT Medical Physics Laboratory, Via Enrico Fermi 15, 01100 Viterbo (Italy)
Institute for Scientific Methodology, Palermo (Italy)
[3] i.e. events corresponding to objective "reductions" of the quantum "state". We adopt here the terminology suggested by Penrose [2,3,4].

*time*. The physical laws which describe the evolution of the wavefunctions on the PSR chronotope (or, rather, on the configuration spaces built from it) are thus expressed in terms of a cosmic time, and this implies the validity of the cosmological principle.

The origin of cosmic time must correspond to a physical singularity (big bang). The inertial frames with respect to which the physical laws are formulated, are not arbitrary, therefore, but constitute a substratum of fundamental observers defined by the big bang.

## 2. "Pure" PSR

In the "pure" Fantappié-Arcidiacono Projective Special Relativity (PSR) theory, physical singularities such as the big bang do not exist [1, 5]. The only "singularities" are of geometric type and are represented by the ordinary local lightcone as well as by the De Sitter horizon, both of which have a past and a future surface (Fig.1).

We consider the geodetic projection of a 5-dimensional hyperspherical chronotope onto a plane tangent to it in the observation pointevent (Fig.2). In the passage from real time to an imaginary time, this becomes the geodetic projection of the De Sitter chronotope, whose result is the PSR chronotope or "Castelnuovo chronotope" [6]. In particular, the circles of the hypersphere consisting of points equidistant from the observation pointevent (tangent point) become constant proper time hyperboloid pairs, one of which is located in the observer's past lightcone and the other, symmetrically, in his future lightcone. The maximum circle of the 5-sphere becomes the Cayley-Klein absolute, with its two surfaces, past and future (Fig.1); these two hypersurfaces constitute the De Sitter horizon.

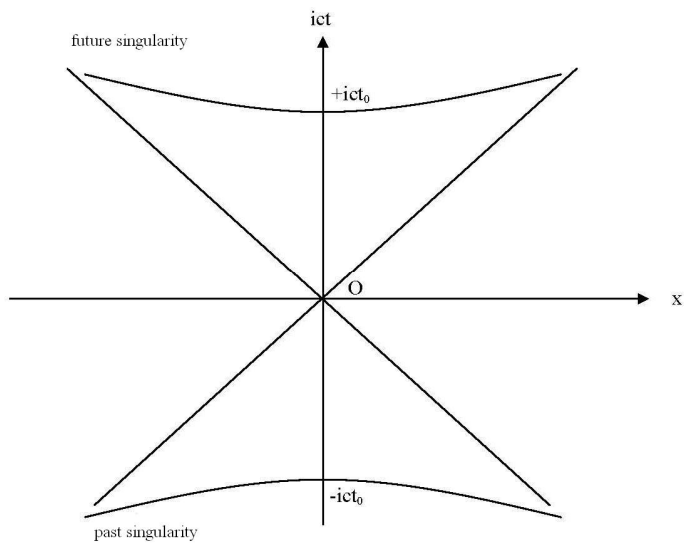

Fig. 1; Castelnuovo chronotope

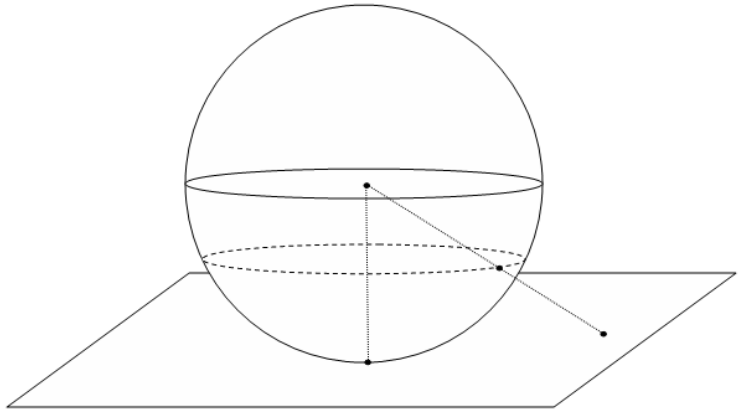

Fig. 2; Geodetic projection of the 5-sphere

In accordance with PSR metrics, the chronological distance of the two surfaces of the absolute from *every* observer equals $t_0$, a fundamental constant of nature. This constant, multiplied by the speed of light in a vacuum *c*, gives the radius *r* of the 5-sphere.
A generic observer sees, in his past lightcone, a Universe in expansion, in which the average motions of the galaxies converge onto the intersection between the universe line of the observer and the past surface of the absolute. This point of intersection, however, is located at the same distance $t_0$ from all observers; thus, the expansion does not occur starting from a true initial physical singularity, but from a geometric horizon. It is therefore a projective relativistic effect connected with the position of the generic observer in the geodetic projection of the De Sitter space, Castelnuovo's spacetime.
As he observes ever more remote galaxies, the observer sees that they tend to thicken indefinitely in an increasingly smaller spacetime zone, concentrated around the past singularity.

## 3. The 5-sphere and the big bang

Because of the cosmological principle, the distance vector ***r***(*t*) of a galaxy from one of the fundamental observers is a function of cosmic time *τ*, according to the relation:

$$\boldsymbol{r}(t) = R(\tau)\boldsymbol{\xi} \tag{1}$$

where $R(\tau)$ is the scale distance (which we shall assume to be dimensionless) and $\boldsymbol{\xi}$ is the position vector of the galaxy in terms of comoving coordinates, independent from *τ*.
The scale distance, which expresses the cosmological model, will have to be derived from the gravitational equations of the Projective General Relativity (PGR), under appropriate initial conditions. In this article we shall assume that $R(0) = 0$, $R(t_0) \sim 1$, in accordance with conventional views on the big bang. The simplest type of imaginable big bang consists of a single pointevent starting from which matter subsequently has expanded in a homogeneous and isotropic manner. This is the traditional image of the big bang, the so-called standard big bang. Of course, more complex hypotheses of a spatially extended big bang cannot be excluded beforehand, but here we shall conform to the simplest hypothesis. We shall see further that a spatially "extended" big bang is actually the most natural choice within 5-sphere geometry.

The Arcidiacono 5-sphere, associated with pure PSR, is described by the equation (in projective coordinates):

$$(x_0)^2 + (x_1)^2 + (x_2)^2 + (x_3)^2 + (x_5)^2 = r^2 \qquad (2)$$

With a Wick rotation of the time coordinate $x_0$, this equation becomes:

$$(x_0)^2 - (x_1)^2 - (x_2)^2 - (x_3)^2 + (x_5)^2 = r^2 \qquad (3)$$

It is therefore a matter of determining the canonical extension of equation (3) that includes equation (1). This extension is obviously:

$$(x_0)^2 - R^2(\tau)[(x_1)^2 + (x_2)^2 + (x_3)^2] + (x_5)^2 = r^2 \qquad (4)$$

As it can easily be seen, by substituting in equation (4) the trivial scale distance $R(\tau) \equiv 1$ we are back to equation (3) and therefore to PSR. There is no real expansion-contraction of space in PSR, because the scale distance is constant. There is, however, a kinematic expansion resulting from the finite value of $r$, in the sense that an observer measuring the ratio between the lengths of rulers remote from him and the lengths of identical rulers near to him finds that it depends on the spatial and chronological distance of the remote rulers.

To provide equation (4) with a clear physical meaning, $\tau$ must be a continuous monotone function of $x_0$; this function will be obtained further on. Let us assume in any case that we have chosen the offset such that $x_0 = 0$ when $\tau = 0$. In this case, the big bang will be represented by the intersection of equation (4) with the hyperplane $x_0 = 0$, i.e., since $R(0) = 0$, by the two points of equation $(x_5)^2 = r^2$. These two points are the points of intersection of the $x_5$ axis with the 5-sphere (2).

The space, as commonly understood, is the three-dimensional hypersurface contained in the hyperplane $x_0 = 0$, perpendicular to the $x_5$ axis, tangent to the hypersphere (2) in the point $x_5 = r$. Thus the choice of the $x_5$ axis, which can be made in $\infty^3$ ways, determines the choice of a three-dimensional space, which is the private space of the fundamental observer adopting the coordinates $x_1, x_2, x_3$. The metrics of this space is described by the second, third and fourth terms of equation (4); it is a space initially constituted by a single point (the fundamental observer himself), which then expands while maintaining its centre in the observer, in accordance with equation (1).

A rotation of the $x_5$ axis in the hyperplane $x_0 = 0$ is equivalent to a constant-time space translation, carried out at the instant of the big bang, which leads from one fundamental observer to another. This result differs radically from the traditional view of the big bang, though it is from there that we started. In the traditional view, the big bang is constituted by a single pointevent in which all the fundamental observers converge. The big bang described by equation (4), on the other hand, is a three-dimensional manifold, in which $\infty^3$ fundamental observers can be selected, each with his own private space in which equation (1) is valid; the coordinates $\xi$ are the comoving coordinates of the galaxy in this space.

At the time $\tau = 0$ there are, therefore, infinite distinct fundamental observers, linked to each other by global coordinate transformations of the De Sitter group. Equation (1) concerns *one* of these observers, as equation (4) does. This result, unknown in Einstein's relativistic cosmology, which, however, we shall come back to, derives from the fact that in this view the big bang does not create space but occurs in a pre-existing (5-dimensional) space.

## 4. The big bang and cosmic time

Let us analyse in detailed way the connection between $\underline{x}_0$ and $\tau$. Let us assume, first of all, that the $\underline{x}_0$ axis is a privileged axis (we shall investigate the nature of this peculiarity at a later stage).
We have hypothesized that the big bang is the set of points of the manifold (4) for which $\underline{x}_0 = 0$. We also require that $\underline{x}_0$ and $\tau$ have to be linked by a continuous monotone function; thus, on manifold (4) the set of observation pointevents corresponding to a same cosmic time will be given by the intersection with the hyperplane $\underline{x}_0$ = costant. The distance of these points from the $\underline{x}_0$ axis is clearly $[r^2 - (\underline{x}_0)^2]^{1/2}$.
Let us therefore consider the plane of contemporaneousness of one of these pointevents, say P. This hyperplane will form an angle $\varphi$ with the big bang hyperplane whose equation is $\underline{x}_0 = 0$. The tangent of this angle will be $\underline{x}_0/[r^2 - (\underline{x}_0)^2]^{1/2}$.
On the four-dimensional hyperplane tangent to equation (4) in P (private chronotope of P) the big bang will thus be projected geodetically at a chronological distance $\tau$ from P such that $\tau/t_0 = \mathrm{tg}(\varphi)$. The sought relation between $\underline{x}_0$ and $\tau$ is therefore:

$$\tau = t_0\, \underline{x}_0/[r^2 - (\underline{x}_0)^2]^{1/2}. \tag{5}$$

It can be noted that $\tau$ increases indefinitely for $\underline{x}_0 \to r$. This limit corresponds to the situation $\varphi \to \pi/2$, in which the geodetic projection of the big bang is crushed onto the De Sitter horizon of the past lightcone of P.
Equation (5) satisfies the previously required condition $\tau = 0$ for $\underline{x}_0 = 0$.

## 5. Redefinition of Milne's "two time scales"

The cosmic time defined by equation (5) is a projective cosmic time; it includes the De Sitter expansion, which is purely kinematic. To move on to a description in which no kinematic component of expansion exists - which is useful, for example, to enable the results of PGR to be compared with those of Einstein's General Relativity (GR) - it is necessary to regraduate cosmic clocks. This regraduation causes a fictitious cosmic repulsion (cosmological term) to appear, which is absent from PGR equations.
In ref. [5] a relation has been found between cosmic time $\tau$ measured in PGR and the corresponding GR cosmic time $\tau'$. This relation is expressed by the "two time scales" already considered by Milne within the framework of Kinematic Relativity (with different meanings of the variables):

$$\tau' = t_0 + t_0 \ln\left(\frac{\tau}{t_0}\right). \tag{6}$$

In equation (6), $\tau$ is defined as equal to zero at the distance $t_0$ from the observer. If, as it is currently customary, an offset is introduced in the definition of $\tau'$ so that this variable is null when $\tau$ equals the chronological distance $t_{BB}$ of the big bang from the De Sitter horizon (a distance which is clearly a function of the cosmic time $\tau$ measured by the observer), the following redefinition is obtained:

$$\tau' \to \tau' = \left[ t_0 + t_0 \ln\left(\frac{\tau}{t_0}\right) \right] - \left[ t_0 + t_0 \ln\left(\frac{t_{BB}(\tau)}{t_0}\right) \right] = t_0 \ln\left(\frac{\tau}{t_0}\right) - t_0 \ln\left(\frac{t_{BB}(\tau)}{t_0}\right). \quad (7)$$

It can easily be seen that:

1) for $\tau = 0$, $\tau' = -\infty$; the De Sitter horizon does not belong to the GR chronotope;

2) for $\tau = t_{BB} = t_0$, i.e. at the big bang, $\tau' = 0$ as required;

3) only the values $\tau \geq t_{BB}(\tau)$ correspond to real physical phenomena;

4) the property $\tau' = t_0$ for $\tau = t_0$, typical of equation (6), returns in the special case $t_{BB}(t_0) = t_0/e$.

The cosmic time $\tau$ which appears in equation (1) is actually $\tau'$. It is the time normally used in astronomical observatories to coordinate events on a cosmological scale.
The effect of the Milne regraduation, equation (7), is to generate a cosmological term $\lambda \approx 1/t_0^2$, and so accelerate the expansion [1].

## 6. The archaic variable $x_0$ as time "precursor"

Let us consider the intersection of (4) with the hyperplane $x_0 = c\theta$, with $c\theta/r \ll 1$. Naturally, the reasoning remains basically the same if one assumes that this intersection, rather than the one with the hyperplane $x_0 = 0$, represents the big bang.
In this case, the intersections of (4) with the hyperplanes $x_0 = c\theta'$, $0 \leq c\theta' \leq c\theta$, are a generalisation of constant cosmic time planes in a subset of the $x_0$ domain where cosmic time is not yet defined, as the big bang "has not yet occurred".
In this sense, it can be said that the archaic variable $x_0$ is a precursor of cosmic time. The subsequent advent of cosmic time conceals the role of $x_0$ which, as can also be seen from equation (5), is the true evolution parameter.

## 7. Big bang - physical aspects. Nucleation

We must now examine what the privileged role of the $x_0$ axis consists in. The big bang is a sort of nucleation of matter and fields by vacuum; to try to understand it we must therefore make some assumptions regarding a pre-cosmic state of matter and energy. The introduction of an evolution parameter such as $x_0$ which can be extended to pre-big bang situations is absolutely necessary if we are to define any pre-cosmic dynamics.

We may assume that the Universe evolves starting from the state $x_0 = 0$, as a complex of fields and particles located on an equator of the 5-sphere (2). The subsequent states are marked by subsequent, increasing values of $x_0$ up to $x_0 = c\theta$. For $x_0 = c\theta$ the phase transition commonly known as the big bang occurs; this is when cosmic time makes its appearance, linked to $x_0$ by equation (5). On the plane $x_0 = c\theta$ the $x_5$ axis can be chosen arbitrarily, thus giving rise to $\infty^3$ possible intersections with manifold (4).

All these intersections constitute the big bang; the big bang thus extends into public spacetime. However, at a chosen date of the $x_5$ axis there is a corresponding selection of a fundamental observer, in whose private space the big bang is pointlike, in accordance with equation (1) and with the initial condition $R(0) = 0$.

What does the Universe at $0 \leq x_0 \leq c\theta$ consist of? It certainly does not consist of real particles nor of real interactions among them, as real interactions (R processes) are events which take place in time, and time does not yet exist. We are therefore naturally led to assume that at this stage matter and fields are present in the form of virtual particles, and that the interactions among these particles are virtual as well. If this hypothesis is accepted as correct, this pre-cosmic state must therefore be described by means of a field quantum theory on the 5-sphere.

The metrics is the Euclidean 5-sphere metrics (2), with real time; that is, a time direction cannot be distinguished from the spatial directions - the parameter $x_0$ is the "precursor" of time. Thus, the equations of motion in this phase are static equations, like the Poisson equation of electrostatics, and there is no actual time evolution, nor is there any propagation of waves or particles. We can therefore speak of a "state" or "condition" of the Universe, but not of any evolution or behaviour on its part.

It can be believed, without too much effort, that the state of matter can still be described by means of macroscopic variables. A set of values of these variables can be produced with many different microstates, and the number of these microstates will define the probability $P$ of the macrostate in question. At this point, an entropy $S$ and a temperature $T$ can be introduced, in purely formal terms, by means of the definitions:

$$S = k \ln P \qquad (8)$$

where $k$ is the customary Boltzmann constant and

$$dS/dF = -1/T \qquad (9)$$

where $F$ is the energy that the system would liberate if all the particles and fields which it is made of become real. By combining the two relations, one has:

$$P = \exp(-F/kT). \qquad (10)$$

We shall assume that the probability $P$ is the greatest possible given the constraints on the system, and that all the variables which describe the macrostate are defined and have uniform values over the entire space (i.e. on the section $x_0 = c\theta'$, $0 \leq c\theta' \leq c\theta$), apart from any fluctuations. Thus, the system is in thermodynamic equilibrium, with a partition function that is compatible with equation (10). This assumption, which is easily justified on the basis of statistical arguments (we choose the most probable condition *a priori*, with small fluctuations given the high number of particles), implies that the Universe is homogeneous and isotropic "before" the big bang and at the big bang, i.e. the cosmological principle. $F$ is the thermodynamic potential pertaining to transitions that are isochoric (the volume $V$ of each section $x_0 = c\theta'$ is finite and does not depend on time because time does not exist) and isothermal ($T$ is defined on each section). Thus, $F$ is the Helmholtz free energy.

Since $x_0$ is an evolution parameter, the variables of state can vary with $x_0$. Since $F$ is constant by definition (as we can define it as the energy *actually* liberated at the "following" big bang), $P$ can depend on $x_0$ only through $T$ [7]. So, generally speaking, $T$ is a function of $x_0$ and we can postulate that $T$ is indeed nothing but $x_0$, converted through the relation:

48$$\underline{x}_0 = \hbar c/kT \qquad (11)$$

in which only fundamental constants appear which, as is known, play the role of conversion factors. The formal temperature $T$ assumes an initial value of infinity on the hypersphere equator, then to decrease as $\underline{x}_0$ increases (and this is the "special role" of the $\underline{x}_0$ axis: basically, it is an axis of *temperatures*) until it reaches the minimum value $\hbar c/kr = E_{dB}/k$ at the extreme $\underline{x}_0 = r$; $E_{dB} = \hbar c/r$ is the de Broglie energy.

Therefore, the big bang occurs at a value of $T$ at which all virtual particles become real and all virtual interactions become real. The big bang is nothing but such transition. The transition depopulates the "virtual" component of the material Universe and, consequently, the subsequent cooling of this component to the de Broglie level does not occur; instead, the history of the real Universe begins. The metrics is now expressed by eq. (4). The R processes start transactions and thus wavefunctions begin to propagate whose arguments are the usual physical coordinates, normalized according to equation (4).

The question now is: at what value of $T$ does the transition occur and why? The problem remains open till now and no possibility can be excluded. A simple answer is obtained by considering that $\underline{x}_0$ is the precursor of time and bearing in mind the role of the fundamental time interval $\theta_0 \sim 10^{-23}$ s (chronon), as hypothesized by one of us in reference [8]. In accordance with this description, the minimum duration of a transaction is of the order of $\theta_0$, and thus the minimum time interval between two R processes which represent the extremes of the transaction must be equal to $\theta_0$. This means that if the Universe has been in existence for a time less than $\theta_0$ (or for a time precursor interval less than $\theta_0$), transactions cannot occur and therefore R processes cannot appear. In other words, a Universe with $\underline{x}_0 < c\theta_0$ must consist only of virtual processes.

The big bang thus consists in the fact that at the value $\underline{x}_0 = c\theta_0$ of the "archaic" variable $\underline{x}_0$ real processes are no longer prohibited, and therefore <u>all the processes and interactions that up to that time were virtual become real</u>. The massive conversion to the real state depopulates the virtual state, which therefore vanishes. All the free energy $F$ is made available in the form of energy and matter in R processes and the "manifest" Universe appears. The connection between R processes is now described by wavefunctions; the coordinates which are the arguments of these wavefunctions are different from those of hypersphere (2) as they satisfy the normalization condition (4). With respect to *these* coordinates, time "flows".

The entire process can be summarized by saying that the Universe undergoes cooling as $\underline{x}_0$ increases, and that at the critical temperature:

$$T_C = \hbar/k\theta_0 \qquad (12)$$

(equal to approximately $10^{13}$ °K; it is the temperature above which hadron stabilization begins) it undergoes a phase transition which leads to the nucleation of matter on the equation space $\underline{x}_0 = c\theta_0$. This nucleation is the big bang and, given the very small value of the ratio $c\theta_0/r \sim 10^{-41}$, it can be said that it is as if the big bang occurs on the equator of the 5-sphere; this confirms the geometric reasoning carried out in the previous paragraph.

Thus, at the time of its appearance in the big bang, the Universe is a system in thermodynamic equilibrium at the temperature $T_C$, homogeneous and isotropic because it is defined by macrovariables which are the same everywhere on the section $\underline{x}_0 = c\theta_0$. All observers exiting from the big bang thus see the Universe in the same way and their motions are therefore - apart from any fluctuations - identical under the action of a global invariance group; that is, the cosmological principle applies, and a cosmic time begins.

It is reasonable to hypothesize that primordial R processes, i.e. those which open the first transactions without closing any, are mutually independent. If this is true, two important consequences ensue:

1) The fluctuations of the density of matter in any two points of the Universe at the very first instants of cosmic time are uncorrelated, since they are due to the localization of real particles by primordial R processes.

2) The depopulation of the pre-cosmic virtual state, as it occurs through independent processes, is entirely similar to the radioactive decay of a substance and therefore proceeds at an exponential rhythm in cosmic time.

From the first consequence it derives that, since the Fourier transform of a Dirac delta function correlation is a flat function, the decomposition of the primordial fluctuations on the celestial sphere into spherical harmonics must give a constant power spectrum on all wavelengths. This spectrum is not directly accessible to observation (our instruments stop at the recombination surface) but can be deduced from the power spectrum of the CMB fluctuations amplified by the plasma. The data collected by the Boomerang collaboration are compatible with this framework [11].

As regards the second point, we can assume that the decay constant is of the order of $\theta_0$. If this is so, the initial density of free energy increases from a zero value to a maximum value (relating to the total emptying of the virtual state) in a time equal to a few "chronons". A part from fluctuations, the final mass-energy density will be the same everywhere and will be equal to the ratio between $F$ (the energy released in the transition) and the volume of the section $\underline{x}_0 = c\theta_0$, which is finite[4]. Thus there is never a singular density value; in other words, in public spacetime <u>the big bang is not truly a singularity</u>.

It must be noted that the contraction resulting from the scale distance operates on the private chronotopes of the individual fundamental observers, not on the public spacetime, which remains unchanged. As one approaches the big bang proceeding backwards in cosmic time, the private contemporaneousness space of each observer contracts in one point; but the uncontracted public space will be identical for all observers, unless a Fantappié-Arcidiacono space translation. Given the initial homogeneity, all the fundamental observers will therefore see the same physical cosmic conditions, <u>although the absence of causal correlations between their respective positions</u>.

Two difficulties with the standard model are worked around in this way, i.e. the justification of the initial homogeneity and the appearance of a singularity[5].

---

[4] This volume is practically that of the equator of the 5-sphere, i.e. $2\pi^2 r^3$.

[5] As a possible justification of eq. (11) we can assume the plane $\underline{x}_0 = c\theta$, $0 \leq c\theta \leq c\theta_0$, to be populated only by the ending events of quantum virtual fluctuations originated at $\theta = 0$. The mean energy $kT$ associated to these fluctuations is then $\approx \hbar/\theta$ because the uncertainty principle. Fluctuations whose duration exceeds $\theta_0$ can be terminated by events which are "R" processes.

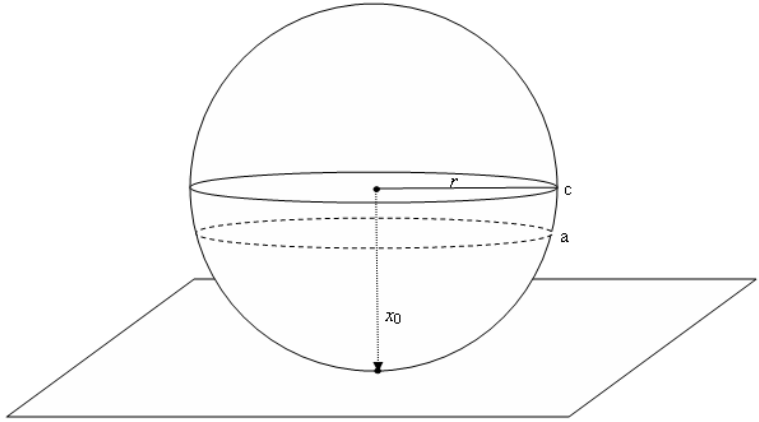

Fig.3; c = equator of the hypersphere [eq. (2)], a = section corresponding to the big bang. The portion of hypersphere between c and a represents the state of the Universe "before" the big bang

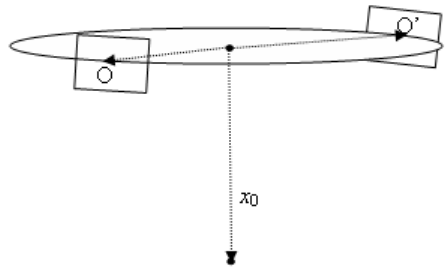

Fig. 4; Different orientations of the fifth axis on the section corresponding to the big bang represent different fundamental observers (O, O', ...). For each of them there is a different flat private spacetime, tangent to the hypersphere in the observer.

## 8. Big bang and microphysics: a Bekenstein relation for the archaic universe

By unifying equations (10) and (11) one has

$$P = \exp(-F\underline{x}_0/\hbar c) = \exp(-p^0\underline{x}_0/\hbar) = \exp(-\Sigma/\hbar) , \qquad (12)$$

where $p^0 = F/c$ and $\Sigma$ is the total action held by the Universe "before" the big bang. It is interesting to note that the following relation exists between the action and the entropy of the pre-big bang Universe:

$$\Sigma/\hbar = -S/k \qquad (13)$$

As it can be seen by direct comparison with equation (8). In other words, $\Sigma$ is a negative entropy or, one might say, a sort of information whose bit is $\hbar\ln(2)$. From equation (12) one has $-\Sigma = \hbar\ln(P)$ and thus, for $P = \frac{1}{2}$ (binary choice), $\Sigma = \hbar\ln(2)$. In general, a dimensionless amount of information $I = \Sigma/[\hbar\ln(2)]$ can be introduced.

From the relation $\underline{x}_0 \leq c\theta_0$, which is valid in the "pre-big bang" era, if one puts $c\theta_0 = 2\pi\underline{R}$ one has $p^0\underline{x}_0 \leq 2\pi p^0\underline{R}$, i.e. $\Sigma \leq 2\pi F\underline{R}/c$. Thus:

$$I \leq 2\pi F\underline{R}/[\hbar c\ln(2)] \qquad (14)$$

and this is a form of the Bekenstein relation which is valid for the "pre-big bang" phase.

Let us now consider primordial R processes, i.e. those which open the first transactions after the big bang but do not close any. From a purely formal point of view, it is as if these processes close a previous transaction, causing the collapse of a wavefunction having constant amplitude over the entire Universe (owing to the homogeneity resulting from the thermodynamic equilibrium) and a phase given by the action $\Sigma$. In this strictly formal sense it is therefore possible to introduce a universal wavefunction entering into the big bang:

$$\Psi = \exp(-i\Sigma/\hbar) . \qquad (15)$$

This wavefunction will "collapse" in the primordial R processes, and the collapse will be completed by the complete transformation of the virtual cosmic state into a real state. Whether or not equation (15) is structured in such a way as to ensure a strong entanglement among pre-cosmic particles is a matter of conjecture at present. The general mechanism here outlined does not in any case require such entanglement, since it is not based on a decoherence process: equation (15) can also be a mere product of wavefunctions of independent particles, which has not been symmetrized in any way.

By comparing equations (12) and (15) it can be seen that $P$ is converted into $\Psi$ if $\Sigma \rightarrow i\Sigma$; this is the "Wick rotation". The Wick rotation has been abundantly discussed in references [1, 7]. We only observe here that, as explained in ref. [1], the Wick rotation also remains in operation after the big bang, in those tiny *little bangs* and *little crunches* which are, respectively, R processes consisting of creation and annihilation of quantum states (quantum leaps).

The framework thus outlined bears a very close resemblance to the Hartle-Hawking solution [7, 12], though it differs from it in a number of aspects, e.g. for its independence from any reasoning in terms of quantum gravity. Equation (12), particularly, shows an inverse exponential dependence of the pre-cosmic state probability from $\underline{x}_0$, which is very similar to a sort of "evanescent wave". This suggests an analogy between the conversion $P \rightarrow \Psi$ and the tunnel effect.

A problem which remains completely open is whether the dominance of matter over antimatter is already present at the big bang or whether it develops starting from a basically balanced situation. Natural symmetries do not appear to emerge which would impose, in this model, an *ab initio* dominance; on the contrary, the hypothesis of maximum initial disorder would seem to support absolute equality. It is possible that the statistical equality was disrupted by small fluctuations, which were subsequently amplified by mechanisms such as those suggested by Sakharov, reaching the present dominance of matter.

## 9. Gravitational equations

We now propose to solve the gravitational equations of Projective General Relativity (PGR) so as to find the correct scale distance $R(\tau)$ as a function of cosmic time $\tau$. By substituting this function in equation (4), the geometry of the model remains completely specified. We shall initially follow the original scheme of the theory proposed by Arcidiacono [6], though, as we shall see, this will have to undergo some adaptation.

In accordance with this scheme, the problem is determined entirely by the two equations:

$$R_{AB} - \frac{1}{2}\check{R}\gamma_{AB} = \chi T_{AB} \qquad (16)$$

$$T_{AB} = \left(\mu + \frac{p}{c^2}\right)\left(\underline{u}_A \underline{u}_B - h^2 \underline{x}_A \underline{x}_B\right) + p\gamma_{AB} \qquad (17)$$

In these equations, $R_{AB}$ is the projective curvature contracted tensor and $\check{R} = \gamma^{AB} R_{AB}$ is its invariant. The energy tensor $T_{AB}$ is defined as a function of the velocity $\underline{u}_A$ of the cosmic fluid in the point $\underline{x}_A$, of its pressure $p$ and of its mass density $\mu$. It has been assumed that $h = 1/t_0$.

The indices $A$, $B$, $C$, ... run along the values 0, 1, 2, 3, 5. We shall use Greek letters such as $\mu$, $\nu$, ... for values along 0, 1, 2, 3; instead, we shall use lowercase Latin letters such as $i$, $j$, ... for values along the spatial coordinates 1, 2, 3. The 0-th coordinate is the time coordinate. $\gamma_{AB}$ is the projective metrics, which we shall proceed to introduce.

### 9.1 Projective metrics

In the PGR scheme, the projective metrics:

$$ds^2 = \gamma_{AB}\, d\underline{x}^A\, d\underline{x}^B \qquad (18)$$

satisfies the two normalization conditions:

$$\gamma_{AB}\, \underline{x}^A\, \underline{x}^B = r^2 \quad , \qquad (19)$$

$$\gamma_{AB} \underline{u}^A \underline{u}^B = -c^2 \quad . \tag{20}$$

In our case, equation (19) is nothing but equation (4); the projective metrics coefficients that are not identically null are therefore identified as follows:

$$\gamma_{0A} = \delta_{0A} \; ; \quad \gamma_{5A} = \delta_{5A} \; ; \quad \gamma_{ij} = -R^2(\tau)\delta_{ij} \quad . \tag{21}$$

By substituting equation (21) into equation (18) one obtains:

$$ds^2 = d\underline{x}_0 \, d\underline{x}^0 - R^2(\tau) d\underline{x}_i \, d\underline{x}^i + d\underline{x}_5 \, d\underline{x}^5 \quad . \tag{22}$$

In equation (22), the function $\tau(\underline{x}_0)$ is given by equation (5). The connection between the projective coordinates $\underline{x}_A$ and the physical ones $x_\mu$ is expressed by the following general relations:

$$\underline{x}_\mu = x_\mu / A \tag{23a}$$

$$\underline{x}_5 = r / A \tag{23b}$$

$$A^2 = \gamma_{55} + 2\gamma_{\mu 5} \frac{x^\mu}{r} + \gamma_{\mu\nu} \frac{x^\mu x^\nu}{r^2} \quad . \tag{23c}$$

In our special case, equation (23c) becomes:

$$A^2 = 1 + x_0 x^0 / r^2 - R^2(\tau) x_i x^i / r^2 \quad . \tag{23d}$$

The projective derivative $\overline{\partial}_A = \partial / \partial \underline{x}_A$ can be expressed as a function of the ordinary one by means of the following relations:

$$\overline{\partial}_\mu = A \partial_\mu \tag{24a}$$

$$\overline{\partial}_5 = -\frac{A}{r} x_\mu \partial^\mu \quad . \tag{24b}$$

For $R(\tau) \equiv 1$ all these relations are converted back to the PSR ones. In the limit $t_0 \to \infty$ one has $A = 1$, and they are transformed back into the ordinary expressions of Einstein's Special Relativity. We note, in particular, that:

$$\tau = \frac{t_0 \underline{x}_0}{\sqrt{r^2 - \underline{x}_0^2}} = \frac{t_0 \frac{x_0}{A}}{\sqrt{r^2 - \left(\frac{x_0}{A}\right)^2}} = \frac{t_0 x_0}{rA\sqrt{1 - \left(\frac{x_0}{rA}\right)^2}} = \frac{x_0}{cA\sqrt{1 - \left(\frac{x_0}{rA}\right)^2}} \quad ;$$

and therefore, in the limit $t_0 \to \infty$ one has $\tau \to x_0/c$. In the same limit, as it can easily be seen by differentiating equation (23b), one has $d\underline{x_5} = 0$. Furthermore, the projective spacetime coordinates are converted into the physical ones, as can be seen from equation (23a). Equation (22) thus becomes:

$$ds^2 = c^2 d\tau^2 - R^2(\tau) dx_i \, dx^i \quad , \tag{25}$$

i.e. ordinary Robertson-Walker metrics with a null spatial curvature (k = 0). To sum up, equation (4) leads to PGR metrics (22) which represents the projective generalization of RW metrics with a Euclidean spatial section.

In the projective metrics (18) the projective coordinates can be substituted by the physical ones, thus obtaining the induced metrics. The induced metric tensor is:

$$g_{\mu\nu} = A^{-4} \left[ A^2 \gamma_{\mu\nu} + (Y_\mu - X_\mu)(Y_\nu - X_\nu) \right] \tag{26a}$$

where:

$$X_\mu = \gamma_{\mu 0} + \gamma_{\mu\nu} x^\nu \quad , \tag{26b}$$

$$Y_\mu = \frac{1}{2} \left[ \partial_\mu \gamma_{00} + x^\nu \partial_\mu \gamma_{\nu 0} + x^\xi x^\nu \partial_\mu \gamma_{\xi\nu} \right] \quad . \tag{26c}$$

In the case of equation (22), one has:

$$X_0 = 1 + x_0; \quad X_i = -R^2(\tau) x_i; \quad Y_0 = -\frac{l^2}{2} \partial_0 (R^2); \quad Y_i = 0. \tag{27}$$

It has been assumed that $l^2 = x_i x^i$. As it can easily be seen, the metrics (26a) is not symmetric; while $g_{0i}$ contains the term

$$R^2 x_i \left( \frac{l^2}{2} \partial_0 (R^2) + 1 + x_0 \right) \quad ,$$

the coefficient $g_{i0}$ contains the term

$$R^2 x_i \left( -\frac{l^2}{2} \partial_0 (R^2) + 1 + x_0 \right) ;$$

so that $g_{i0} \neq g_{0i}$. However, near the observer (i.e. in the local limit $l \to 0$) the metrics induced in the physical coordinates is symmetric.

## 9. 2 Projective connection

We now have all the elements for determining the left-hand member of equation (16). The first step is to calculate the projective connection coefficients:

$$\pi^A_{BC} = \frac{1}{2}\gamma^{AS}\left(\bar{\partial}_B \gamma_{CS} + \bar{\partial}_C \gamma_{BS} - \bar{\partial}_S \gamma_{BC}\right) , \qquad (28)$$

where

$$\gamma^{AS} \gamma_{BS} = \delta^A_B . \qquad (29)$$

In the specific case of metrics (21) the only non-null coefficients are the following:

$$\pi^i_{0i} = \pi^i_{i0} = \frac{R'}{R}\frac{d\tau}{d\underline{x}_0}$$

$$\pi^0_{ii} = RR'\frac{d\tau}{d\underline{x}_0} . \qquad (30)$$

In equations (30), the prime indicates the derivation with respect to $\tau$ and:

$$\frac{d\tau}{d\underline{x}_0} = \frac{1}{c}\left[1+\left(\frac{\tau}{t_0}\right)^2\right]^{3/2} , \qquad (31)$$

as it can be seen by deriving equation (5) with respect to $\underline{x}_0$ and then substituting equation (5) into the expression obtained. If cosmic time is conventionally measured as a length (i.e. indicating the product $c\tau$ with the symbol $\tau$), the factor $1/c$ in the second member of equation (31) disappears. In this case, in Einstein's limit $t_0 \to \infty$ equation (31) becomes:

$$\frac{d\tau}{d\underline{x}_0} = 1 . \qquad (32)$$

By further deriving equation (31) one obtains, with the same convention ($t_0$ is actually $r$):

$$\frac{d^2\tau}{d\underline{x}_0^2} = \frac{3\tau}{t_0^2}\left[1+\left(\frac{\tau}{t_0}\right)^2\right]^2 . \qquad (33)$$

In Einstein's limit $t_0 \to \infty$ this expression becomes:

$$\frac{d^2\tau}{d\underline{x}_0^2} = 0 . \qquad (34)$$

Equations (32), (34) will come in useful later.

### 9.3 Projective curvature

The next step is the calculation of the projective curvature contracted tensor:

$$R_{AB} = \bar{\partial}_A \pi^C_{BC} - \bar{\partial}_C \pi^C_{BA} + \pi^C_{BD} \pi^D_{CA} - \pi^C_{BA} \pi^D_{CD} \quad . \tag{35}$$

Through some tedious but consequential passages, it can be seen that in the case of metrics (21) the only non-null components of this tensor (which, one must bear in mind, also includes torsion [6]) are:

$$R_{00} = 3\left(\frac{R'}{R}\right)\left(\frac{d^2\tau}{dx_0^2}\right) + 3\left(\frac{R''}{R}\right)\left(\frac{d\tau}{dx_0}\right)^2 \quad ;$$

$$R_{11} = R_{22} = R_{33} = -2(R')^2\left(\frac{d\tau}{dx_0}\right)^2 - RR''\left(\frac{d\tau}{dx_0}\right)^2 - RR'\left(\frac{d^2\tau}{dx_0^2}\right) \quad .$$

(36)

In the limit $t_0 \to \infty$ these expressions become:

$$R_{00} = 3\frac{R''}{R} \quad ; \qquad R_{11} = R_{22} = R_{33} = -2(R')^2 - RR'' \quad ; \tag{37}$$

As it can easily be seen by substituting equations (32), (34).
The tensor invariant becomes, in the case of metrics (21):

$$\check{R} = \gamma^{AB} R_{AB} = R_{00} - \frac{3}{R^2} R_{11} \quad . \tag{38}$$

### 9.4 Definition of the energy tensor

Let us now turn our attention to the right-hand member of equation (16). Equation (17) represents the energy tensor of an arbitrary fluid; it must be specialized for the particular case of cosmic fluid. The fluid velocity field $u_A$ will be a function of the position and time and the structure of this function will be peculiar to cosmic fluid.
Cosmic fluid will be defined by the property which locally, i.e. in the observation pointevent $x_A = 0$, its velocity is oriented along the $x_0$ axis, which we have chosen as the cosmic time axis. Also, the pressure and density of the fluid, because of the cosmological principle, will depend solely on the cosmic time $\tau$.
Let us now see how to express these requirements formally. Firstly, it follows from equation (21) and from equation (29) that

$$\gamma^{0A} = \delta^{0A} \; ; \quad \gamma^{5A} = \delta^{5A} \; ; \quad \gamma^{ij} = -\delta^{ij}/R^2(\tau) \; . \tag{39}$$

Thus:

$$\underline{u}_A \underline{u}^A = \underline{u}_A \gamma^{AB} \underline{u}_B = (\underline{u}_0)^2 - \frac{1}{R^2}[(\underline{u}_1)^2 + (\underline{u}_2)^2 + (\underline{u}_3)^2] + (\underline{u}_5)^2 \; . \tag{40}$$

$$\underline{u}_A \underline{x}^A = \underline{u}_A \gamma^{AB} \underline{x}_B = (\underline{u}_0 \underline{x}_0) - \frac{1}{R^2}[(\underline{u}_1 \underline{x}_1) + (\underline{u}_2 \underline{x}_2) + (\underline{u}_3 \underline{x}_3)] + (\underline{u}_5 \underline{x}_5) \; . \tag{41}$$

Let us consider the conditions valid for a generic fluid [6, 9]:

$$\underline{u}_A \underline{u}^A = -c^2 \; ; \qquad \underline{u}_A \underline{x}^A = 0 \; . \tag{42}$$

Since the fluid velocity is parallel to $\underline{x}_0$, its spatial components must be null. Equations (42) thus take the form:

$$(\underline{u}_0)^2 + (\underline{u}_5)^2 = -c^2 \; ; \qquad (\underline{u}_0 \underline{x}_0) + (\underline{u}_5 \underline{x}_5) = 0 \; .$$

It follows from the second relation that:

$$\underline{u}_5 = -\underline{u}_0 \underline{x}_0/\underline{x}_5 \; ,$$

and by substituting this result into the first relation one obtains:

$$\underline{u}_0 = \frac{ic}{\sqrt{1+\left(\dfrac{\underline{x}_0}{\underline{x}_5}\right)^2}} \; . \tag{43}$$

The ambiguity regarding the sign of $\underline{u}_0$ is removed if it is observed that the direction of $\underline{u}_0$ is towards the future. From the second relation one thus obtains:

$$\underline{u}_5 = \frac{-ic}{\sqrt{1+\left(\dfrac{\underline{x}_0}{\underline{x}_5}\right)^2}} \left(\dfrac{\underline{x}_0}{\underline{x}_5}\right) . \tag{44}$$

As it can be seen, $\underline{u}_0$ and $\underline{u}_5$ are determined as functions of the coordinates, in that $\underline{x}_0/\underline{x}_5 = x_0/r$. The components of the energy tensor become, therefore:

$$T_{00} = \left(\mu + \frac{p}{c^2}\right)\left[(\underline{u}_0)^2 - h^2(\underline{x}_0)^2\right] + p$$

$$T_{0i} = -\left(\mu + \frac{p}{c^2}\right)h^2 \underline{x}_0 \underline{x}_i$$

$$T_{ij} = -\left(\mu + \frac{p}{c^2}\right)h^2 \underline{x}_0 \underline{x}_i - pR^2\delta_{ij}$$

$$T_{55} = \left(\mu + \frac{p}{c^2}\right)\left[(\underline{u}_5)^2 - h^2(\underline{x}_5)^2\right] + p$$

$$T_{5i} = -\left(\mu + \frac{p}{c^2}\right)h^2 \underline{x}_5 \underline{x}_i$$

$$T_{50} = \left(\mu + \frac{p}{c^2}\right)\left(\underline{u}_5 \underline{u}_0 - h^2 \underline{x}_5 \underline{x}_0\right)$$

$$p = p(\tau); \quad \mu = \mu(\tau) \ .$$

(45)

It is interesting to analyse the form taken by this tensor in Einstein's limit $t_0 \to \infty$. Firstly, $h$ cancels out, thus equations (45) become:

$$T_{00} = \left(\mu + \frac{p}{c^2}\right)(\underline{u}_0)^2 + p$$

$$T_{0i} = 0$$

$$T_{ij} = -pR^2\delta_{ij}$$

$$T_{55} = \left(\mu + \frac{p}{c^2}\right)(\underline{u}_5)^2 + p$$

$$T_{5i} = 0$$

$$T_{50} = \left(\mu + \frac{p}{c^2}\right)\underline{u}_5 \underline{u}_0 \ .$$

From equations (43), (44), taking into account that $\underline{x}_0/\underline{x}_5 = x_0/r \to 0$, one has $\underline{u}_0 = ic$ and $\underline{u}_5 = 0$. The only non-null components of the energy tensor are, therefore, in this limit:

$$T_{00} = -\left(\mu + \frac{p}{c^2}\right)c^2 + p = -(\mu c^2 + p) + p = -\mu c^2$$

$$T_{ij} = -pR^2\delta_{ij}; \qquad T_{55} = p \ . \qquad (46)$$

### 9.5 Problems with the Arcidiacono scheme

At this point the expressions (36), (38), (45) would have to be substituted into equation (16) and the thus obtained equations solved. It is easily verified, however, that this is impossible. The left member of equation (16) depends solely on $\tau$, while the right member is also a function of the projective coordinates; a solution $R(\tau)$ that is a function only of $\tau$ cannot therefore exist. Furthermore, the system is overdetermined, because there are six distinct equations for the various non-null components of the energy tensor (45), but only three unknown quantities $p(\tau)$, $\mu(\tau)$, $R(\tau)$. Apart from the fact that the equation of state would become redundant (which is in itself absurd), it is easily verified that the obtained system is incompatible.

One may wonder whether equations (16) are incompatible with the cosmological principle or whether they are inconsistent in general. A simple reflection on the physical foundations clearly shows the general inconsistence of equations (16) and indicates how to work around this problem.

Let us consider the emission of a physical signal (wave or particle) in a pointevent A and its subsequent detection in a pointevent B. Emission in A and detection in B are *local interactions*. Given the local nature of these phenomena, the presence/absence of the cosmological element represented by the De Sitter horizon cannot modify their structure[6]. Instead, this element will modify the *propagation* of the signal from A to B, in the sense that, once the characteristics of signal in A have been established, the characteristics detected in B will be different according to whether or not a De Sitter horizon exists. Thus, in PSR (PGR), the laws of signal propagation are altered, with respect to what is stated in SR (GR), while the description of the interactions remains unchanged.

The projective coordinates and their metrics (18) describe, therefore, the projective effects on the propagation of signals arriving at the observer placed in the pointevent $\underline{x}_\mu = 0$; however, the local physics, i.e. the structure of the interaction events which occur in $\underline{x}_\mu = 0$, is not influenced by these effects. Now, equations (16) must indeed describe the structure of an interaction, the gravitational interaction, which is local. What equations (16) must describe is the coupling of the local matter-energy distribution with the local spacetime curvature-torsion. It can be observed, however, that this is not so. Equations (45) describe the local matter-energy distribution only in Einstein's limit $t_0 \to \infty$, when they are converted into equations (46). Similarly, the contracted projective curvature tensor (35) describes the local curvature-torsion only in the limit $t_0 \to \infty$. In this limit, the projective coordinates once again become the customary physical coordinates (which identify the *interaction events*) and the projective effects disappear.

The Arcidiacono gravitational theory is therefore physically inconsistent in general, and not only in the particular case of the cosmological problem discussed here. Before proceeding further, a general redefinition of the Arcidiacono scheme is therefore necessary.

---

[6] For example, there can be no projective effect that induces a cosmological variation of the interaction constants, as some, instead, have speculated.

**9.6 Revision of the PGR scheme**

We propose to reformulate PGR through the following three axioms.

1) The most general expression of the metrics (18) is the following:

$$ds^2 = \gamma_{\mu\nu}(\underline{x})d\underline{x}^\mu d\underline{x}^\nu + d\underline{x}_5 d\underline{x}^5 \ . \tag{47}$$

The $\gamma_{\mu\nu}$ must be determined for each specific case. This is provided for by the following two axioms.

2) Gravitational equations. We shall assume that they are the Arcidiacono ones, but considered in the local limit $t_0 \to \infty$:

$$\lim_{t_0 \to \infty} (R_{AB} - \frac{1}{2}\breve{R}\gamma_{AB} - \chi T_{AB}) = 0 \ . \tag{48}$$

The limit operation causes the disappearance of the components with $A$ and/or $B$ equal to 5, and reduces the system to Einstein's ordinary equations, whose solution is the tensor $\gamma_{\mu\nu}(x)$. However, the expression in parentheses admits the De Sitter group as a holonomy group, rather than the Poincaré group of ordinary GR. In other words, the covariance of equations (48) is of a different type from that of the conventional Einstein equations. Also, the normalization condition (19) applies here, which, by virtue of equation (47), takes the form:

$$\gamma_{\mu\nu}(\underline{x})\underline{x}^\mu \underline{x}^\nu + \underline{x}_5 \underline{x}^5 = r^2 \ . \tag{49}$$

Equation (49) can be considered as the widest physically acceptable generalization of equation (3). These two facts are unknown in ordinary GR and it is precisely these which, in the special case of the cosmological problem discussed here, lead to the freedom of choice of the fifth axis at the instant of the big bang and, therefore, to the existence of a plurality of distinct fundamental observers for $\tau = 0$.

3) Projective prolongation of the metrics. Let us postulate that the tensor $\gamma_{\mu\nu}(\underline{x})$ which appears in equations (47), (49) can be deduced from the $\gamma_{\mu\nu}(x)$, which are solutions of equation (48), through the substitution of the physical coordinates with the corresponding projective coordinates [eqs. (23a)]. A consequence of this postulate is that $\gamma_{\mu\nu}(\underline{x})$ does not depend explicitly on $\underline{x}_5$.
This axiom modifies the metrics derived from the local physics (interactions), including in it the projective effects induced by the existence of a De Sitter horizon.

If, for example, one wished to treat the problem of the motion of Mercury with this scheme, it would firstly be necessary to resolve the ordinary GR equations; then, in the thus obtained metric tensor, the spacetime coordinates would have to be substituted by the projective ones, using eqs. (23a). This tensor should be substituted in equations (47), (49). The metrics induced in physical coordinates by equation (47) would then include the projective effects on motion.

## 10. Determination of the scale distance

Let us now return to the cosmological problem. We shall no longer apply the original PGR scheme to it, but the modified one described in the previous section. How equation (48) prescribes, we must work with the curvature tensor and energy tensor expressions valid in Einstein's limit. Furthermore, the component 55 of the energy tensor no longer bears any weight, because a gravitational equation for it no longer exists. We thus have, from equations (37), (38):

$$\breve{R} = 6\frac{R''}{R} + 6\left(\frac{R'}{R}\right)^2 ,$$

$$G_{00} = R_{00} - \frac{1}{2}\breve{R}\gamma_{00} = -3\left(\frac{R'}{R}\right)^2 ,$$

$$G_{ii} = R_{ii} - \frac{1}{2}\breve{R}\gamma_{ii} = (R')^2 + 2RR'' .$$

Bearing in mind equations (46), equations (48) become:

$$-3\left(\frac{R'}{R}\right)^2 = -\chi\mu c^2$$

$$(R')^2 + 2RR'' = \chi(-pR^2) .$$

These relations can then be rearranged in the customary form:

$$\left(\frac{R'}{R}\right)^2 = \frac{\chi\mu c^2}{3} \tag{50}$$

$$\left(\frac{R'}{R}\right)^2 + 2\left(\frac{R''}{R}\right) = -\chi p \tag{51}$$

which is that of the Fridman cosmological model having the spatial curvature index k = 0. As it can be seen, unlike as in conventional GR, <u>the spatial section is necessarily flat here</u>, as an effect of equations (22), (25). Consequently, there is no critical value of $\mu$ at which space is flat. This removes the flatness problem afflicting the standard model.

## 11. Kinematic origin of the cosmological term

There is general agreement on the existence of a cosmological term. This term does not appear in equations (48), and this means that no true repulsive force nor any "vacuum energy" exist. As we touched on in the first part, the cosmological term appears as a consequence of the regraduation of cosmic clocks [eqs. (6), (7)], which we rewrite here in short form:

$$\tau' = \ln\left(\frac{\tau}{t_0}\right) + k(\tau) \quad . \tag{52}$$

Of course, the passage from $\tau$ to $\tau'$ does not alter the structure either of equation (22) or of equation (25). It must be remembered that for a Universe made of disgregated matter (dust) one has $p = 0$ and the general solution of equations (50), (51) is

$$R(\tau) = R_0 \, \tau^{2/3} \quad . \tag{53}$$

That equation (53) is the solution to equations (50), (51) can easily be verified by direct substitution. One has, among other things, that:

$$\chi\mu c^2 = \frac{4}{3}\tau^{-2} = \frac{4}{3}t_0^{-2}\exp\left[-2\left(\frac{\tau'-k}{t_0}\right)\right] = \frac{4}{3}t_0^{-2}\exp\left[2\left(\frac{k-\tau'}{t_0}\right)\right] \quad ,$$

in which equation (52) has been inserted. By inverting this relation we find:

$$k(\tau') = \tau' + \frac{t_0}{2}\ln\left(\frac{3t_0^2}{4}\right) + \frac{t_0}{2}\ln(\chi\mu c^2) \quad . \tag{54}$$

From here onwards we shall indicate with a dot the derivation with respect to $\tau'$, to distinguish from the derivation with respect to $\tau$, which will be marked with a prime. Deriving equation (54) with respect to $\tau'$ one has:

$$\dot{k}(\tau) = 1 + \frac{t_0}{2}\frac{\dot{\mu}}{\mu} \quad . \tag{55}$$

By inverting equation (52) one has:

$$\tau = t_0 \exp\left(\frac{\tau'-k}{t_0}\right) \quad . \tag{56}$$

Inserting equation (56) into equation (53) one obtains, apart from a non-essential redefinition of $R_0$:

$$R = R_0 \exp\left[\tau'\sqrt{\frac{\lambda}{3}} - \frac{2}{3}\frac{k(\tau')}{t_0}\right] \quad , \tag{57}$$

where we have let $\lambda = 12/(9t_0^2)$. Therefore:

$$\dot{R} = \left(\sqrt{\frac{\lambda}{3}} - \frac{2\dot{k}}{3t_0}\right) R \quad , \tag{58a}$$

$$\ddot{R} = -\frac{2}{3t_0}\ddot{k}R + \left(\sqrt{\frac{\lambda}{3}} - \frac{2\dot{k}}{3t_0}\right)^2 R \quad . \tag{58b}$$

It follows from equation (58a) that:

$$\left(\frac{\dot{R}}{R}\right)^2 = \left(\sqrt{\frac{\lambda}{3}} - \frac{2\dot{k}}{3t_0}\right)^2 = \frac{\lambda}{3} + \frac{4\dot{k}^2}{9t_0^2} - \frac{4\dot{k}}{3t_0}\sqrt{\frac{\lambda}{3}} \quad .$$

Let us suppose, then, that:

$$\chi\mu c^2 = \frac{4\dot{k}^2}{3t_0^2} - \frac{4\dot{k}}{t_0}\sqrt{\frac{\lambda}{3}} = \frac{4}{3t_0^2}(\dot{k}^2 - 2\dot{k}) \quad ; \tag{59}$$

In this case, the last relation becomes:

$$\left(\frac{\dot{R}}{R}\right)^2 = \frac{\lambda}{3} + \frac{\chi\mu c^2}{3} \quad . \tag{60}$$

In the same hypothesis, equation (58b) becomes:

$$\frac{\ddot{R}}{R} = -\frac{2}{3t_0}\ddot{k} + \frac{\lambda}{3} + \frac{\chi\mu c^2}{3} \quad .$$

Therefore:

$$2\frac{\ddot{R}}{R} + \left(\frac{\dot{R}}{R}\right)^2 = \lambda + [\chi\mu c^2 - \frac{4\ddot{k}}{3t_0}] \quad .$$

If we now suppose that:

$$\ddot{k} = \frac{3t_0}{4}\chi\mu c^2 \quad , \tag{61}$$

one obtains:

$$2\frac{\ddot{R}}{R} + \left(\frac{\dot{R}}{R}\right)^2 = \lambda \quad . \tag{62}$$

Equations (60), (62) are the Fridman equations for the model $k = 0$, $\lambda = 12/(9t_0^2) > 0$. One therefore has a purely kinematic genesis of the cosmological term, resulting from the regraduation of the cosmic clocks of the fundamental observers. This simply means that the physics essential to the cosmological model is linked to the global geometry of the 5-sphere.

The intermediate assumptions (59) and (61) remain to be justified. Deriving equation (59) we have:

$$\chi\dot{\mu}c^2 = \frac{8\ddot{k}}{3t_0^2}(\dot{k}-1) \ .$$

If equation (61) is inserted into this expression, it becomes:

$$\chi\dot{\mu}c^2 = [\frac{8}{3t_0^2}(\dot{k}-1)][\frac{3t_0}{4}\chi\mu c^2] \quad ,$$

i.e.:

$$\frac{\dot{\mu}}{\mu} = \frac{2}{t_0}(\dot{k}-1) \ .$$

But this result is certainly true, because it coincides with equation (55). Thus the logical conjunction of the two propositions (59) and (61) produces a true proposition, and this is only possible if equation (59) and equation (61) are separately true.

**12. Red shift law and luminosity-distance relation**

The equation of null-length geodetics, which describe light ray propagation, is $A^2 = 1$, i.e.:

$$(x_0)^2 - R^2(\tau')(x_i x^i) = 0.$$

For infinitesimal tracts of the geodetic, this expression becomes:

$$(dx_0)^2 - R^2(\tau')(dx_i dx^i) = 0 \ ,$$

and letting $dl^2 = dx_i dx^i$ one has:

$$l = \int_{(x_0)_{emission}}^{(x_0)_{arrival}} \frac{dx_0}{R(\tau')} \ . \tag{63}$$

Let us now denote with $(\Delta x_0)_{arrival}$ and $(\Delta x_0)_{emission}$ the time interval between two consecutive waves of the same spectral line measured, respectively, at arrival and at emission. A crucial point is that these intervals are measured locally, and therefore are not affected by projective effects. Since their values are very small compared to the integration interval, one has:

$$l = \int_{(x_0)_{emission} + (\Delta x_0)_{emission}}^{(x_0)_{arrival} + (\Delta x_0)_{arrival}} \frac{dx_0}{R(\tau')} = \int_{(x_0)_{emission}}^{(x_0)_{arrival}} \frac{dx_0}{R(\tau')} + \frac{(\Delta x_0)_{arrival}}{R_{arrival}} - \frac{(\Delta x_0)_{emission}}{R_{emission}},$$

and therefore:

$$\frac{(\Delta x_0)_{arrival}}{R_{arrival}} - \frac{(\Delta x_0)_{emission}}{R_{emission}} = 0.$$

But the relation between wavelengths at arrival and at emission is given precisely by:

$$\frac{\lambda_{arrival}}{\lambda_{emission}} = \frac{(\Delta x_0)_{arrival}}{(\Delta x_0)_{emission}},$$

and therefore

$$z = \frac{\lambda_{arrival} - \lambda_{emission}}{\lambda_{emission}} = \frac{(\Delta x_0)_{arrival} - (\Delta x_0)_{emission}}{(\Delta x_0)_{emission}} = \frac{R_{arrival}}{R_{emission}} - 1.$$

That is:

$$1 + z = \frac{R_{arrival}}{R_{emission}}. \tag{64}$$

The law which expresses red shift as a function of the scale distance variation is therefore the same as in ordinary General Relativity. Consequently, also the relation between apparent luminosity **L** and absolute luminosity $\mathbf{L_0}$ remains the same as in General Relativity:

$$\mathbf{L} = \frac{\mathbf{L_0}}{4\pi R_{arrival}^2 \, l^2 \, (1+z)^2}. \tag{65}$$

However, equation (65) does not include the projective effects on the Poynting vector, which cannot be discussed here. We point out the interested reader to ref. [5], where these effects, which are very small, are described in PSR approximation.
Generally speaking, it will be necessary to write Maxwell's electromagnetic equations, in the form generalized by Arcidiacono, in such a way that they be covariant with respect to the metrics (22), and to study the Poynting vector associated with stellar emission as a function of the distance from the source.

The content of this section shows that, apart from small effects on equation (65) which remain to be investigated, the predictive content of the Fridman model $k = 0$, $\lambda > 0$ found in the previous section is the same as in conventional GR.

The model admits, as general solution [10]:

$$R^3(\tau') = \frac{4\pi G \rho_0}{\lambda}\left[\cosh\left(\tau'\sqrt{3\lambda}\right) - 1\right], \tag{66}$$

where $G$ is the Newton's gravitational constant and $\rho_0$ denotes the mass density evaluated when $R = 1$. The Hubble parameter is expressed as:

$$H(\tau') = \frac{8\pi G \rho_0}{3R^3} + \frac{\lambda}{3}. \tag{67}$$

Supposing the actual cosmic time to be $1.29 t_0$, which is acceptable, the variable $\Omega_\Lambda = \lambda/3H_0^2$ equates 0.738, how can be easily checked by substituting eqs. (66), (67). Under the same circumstance, $H_0 = 0.3 \times 10^{-17}$ h s$^{-1}$.

The figure 0.738 is very close to the accepted value 0.74. Because of the flatness, no contribution to $\Omega$ relating to curvature really exists. If the constraint $\Omega = 1$ (which really does not hold) is assumed, one is forced to postulate a contribution to $\Omega$ relating to matter of $1 - 0.74 = 0.26$. The great part of this contribution should be attributed to "dark matter".

**Conclusions**

We have developed a cosmological model based on a Euclidean timeless spatial structure (the 5-sphere) as a substratum of the ordinary metrics used to describe the propagation of wavefunctions. In the proposed geometry, the time variable represents not a local time but a *cosmic time*; this not only justifies the assumption of a cosmological principle and identifies the De Sitter observers class, but has another significant physical meaning. Through the Wick rotation and the emergence of time's arrow, what we propose here is actually the transition from a non-local real-time phase to a local imaginary-time one, and therefore a quantum interpretation of the big bang.

In this interpretation, the pointlike singularity is replaced by a process of nucleation extended upon the entire space $\underline{x}_0 = c\theta_0$, which practically coincides with the 5-sphere equator. At the temperature $T_C = \hbar/k\theta_0$ (equal to approximately $10^{13}$ °K) one has a phase transition which produces a "flocculation" of matter on this space. At the time of its appearance in the big bang, the Universe is a system in thermodynamic equilibrium, homogeneous and isotropic because it is defined by macrovariables which are the same everywhere on the section $\underline{x}_0 = c\theta_0$. All the observers thus see the Universe in the same way and their motions are - apart from any fluctuations - identical under the action of a global invariance group; that is, the cosmological principle applies, and a cosmic time begins.

It is important to note that in this context the term "vacuum" has a very precise meaning, and refers to a pre-cosmic state $\underline{x}_0 < c\theta_0$ of the Universe, consisting only of virtual processes. The big bang consists in the depopulation of the pre-cosmic virtual state, which proceeds in a completely similar manner to radioactive decay at an exponential rate in cosmic time. At the value $\underline{x}_0 = c\theta_0$ of the "archaic" variable $x_0$, real processes are no longer prohibited, and all the processes and interactions that were virtual up to that time become real. For these processes which cause the collapse of the universal wave function $\Psi = \exp(-i\Sigma/\hbar)$, we have adopted the denomination of primordial R processes. By introducing a minimum interval for an elementary quantum transaction (chronon), we easily obtain a Bekenstein relation for the archaic phase of the "virtual" universe. Finally, we have proposed a complete and consistent solution to the Projective General Relativity (PGR) equations, which univocally defines the relation between the scale factor $R(\tau)$ and cosmic time $\tau$. Unlike GR, it is possible to prove that the spatial section is necessarily flat, and therefore the problem of the "flatness" of the standard model is removed. The role of the cosmological term is also different - it appears here as a fictitious cosmic repulsion resulting from a regraduation of the cosmic clocks adopted by fundamental observers.

We believe that the adoption of the physical model for the Euclidean 5-sphere proposed here is the most natural one for developing a quantum cosmology that is consistent and free from the ambiguities of the classical model resulting from the forced cohabitation of the original GR scenario with quantum elements. Its strength lies in the powerful constraints placed by global geometry on the quantum description, but it will be necessary in future to develop in detail a theory of fundamental interactions on the 5-sphere, in order to reach a complete unified description of constraints and processes.